\documentclass[aps,prl,reprint,twocolumn,showpacs,superscriptaddress,amsmath,amssymb]{revtex4-1}

\usepackage{graphicx}  % needed for figures5
\usepackage{dcolumn}   % needed for some tables
\usepackage{bm}        % for math
\usepackage{hyperref}
\usepackage{verbatim}  % for comment
\usepackage{natbib}
\usepackage{epsfig}
\usepackage{epstopdf}
\usepackage{booktabs}  % for multicolumn 
\usepackage{color}
\usepackage[toc,page]{appendix}
\usepackage{placeins}
\usepackage{float}
\usepackage{multirow}
\usepackage[normalem]{ulem}
\usepackage{amsmath}
\usepackage{tabu}
\usepackage[utf8]{inputenc}
\usepackage{rotating}
\usepackage{multirow}
\usepackage[export]{adjustbox}
\usepackage{subcaption}
\usepackage[justification=RaggedRight, singlelinecheck=false]{caption}

\begin{document}

\title{On The Critical Casimir Interaction Between Anisotropic Inclusions On A Membrane}

\author{Jorge Benet}
\affiliation{Department of Physics, Durham University, South Road, Durham DH1 3LE, UK}
\author{Fabien Paillusson}
\email{fpaillusson@lincoln.ac.uk}
\affiliation{School of Mathematics and Physics, University of Lincoln, Lincoln, LN6 7TS, UK}
\author{Halim Kusumaatmaja}
\email{halim.kusumaatmaja@durham.ac.uk}
\affiliation{Department of Physics, Durham University, South Road, Durham DH1 3LE, UK}
 
\date{\today}

\begin{abstract}
Using a lattice model and a versatile thermodynamic integration scheme, we study the critical Casimir interactions between inclusions embedded in a two-dimensional critical binary mixtures. For single-domain inclusions we demonstrate that the interactions are very long range, and their magnitudes strongly depend on the affinity of the inclusions with the species in the binary mixtures, ranging from repulsive when two inclusions have opposing affinities to attractive when they have the same affinities. When one of the inclusions has no preference for either of the species, we find negligible critical Casimir interactions. For multiple-domain inclusions, mimicking the observations that membrane proteins often have several domains with varying affinities to the surrounding lipid species, the presence of domains with opposing affinities does not cancel the interactions altogether. Instead we can observe both attractive and repulsive interactions depending on their relative orientations. With increasing number of domains per inclusion, the range and magnitude of the effective interactions decrease in a similar fashion to those of electrostatic multipoles. Finally, clusters formed by multiple-domain inclusions can result in an effective affinity patterning due to the anisotropic character of the Casimir interactions between the building blocks.
\end{abstract}

\maketitle

\section{Introduction}
Originally, the cell membrane was considered just as a physical barrier that kept cell components together. However, in the last decades, advances in experimental techniques, including atomic force and fluorescence microscopies, have enabled the probing of membranes' inner structure and composition to such extents that it has been necessary to rethink our understanding of its physics, chemistry and its role in biology \cite{shan15_ChemSocRev}. Far from being a simple barrier, cell membranes are in fact very complex mixtures in which lipids and proteins meet, interact and self-assemble. A number of studies have shown that lipids are organized in the lateral dimension \cite{kusumi96_COCB} and that most membrane proteins organize in clusters \cite{uhles03_JCB,kai06_JBiochem,low06_MBoC,sieber06_BioJ}. Further, there is now a growing consensus that such lateral organization and compartmentalisation play important roles in biological processes, such as in cell signalling and membrane trafficking \cite{alonso01_JCellSci,ikonen08_NRMCB,MartinezOutschoorn15_NatRevCancer}; and it has also been suggested to be involved in a number of diseases from HIV \cite{simons02_JClinInv} to liver \cite{holthuis14_Nature} and prion diseases like Alzheimer \cite{critchley04_BBRC}.

Biologically-motivated scenarios, such as the {\it fence and pickets} model \cite{kusumi05_AnnRevBioStr} or the {\it lipid raft} hypothesis \cite{simons97_Nature}, have been put forward to explain lateral membrane organizations. However, to date it remains unclear whether such scenarios are feasible from a chemical physics standpoint and, if possible, to what extent they are harnessed by biological processes {\it in vivo}. In this context, our aim here is to contribute towards understanding how inclusions ({\it e.g.} membrane proteins) may assemble into clusters within a cell membrane. Assuming the aggregation process occurs in thermodynamic equilibrium conditions, two possibilities come to mind. Firstly, it may occur by the free diffusion, collision and irreversible chemical bonding of the inclusions in a process reminiscent of a Diffusion Limited Aggregation \cite{Saxton92, Saxton93, Tian16}. Secondly, the inclusions may instead self-organise into metastable clusters, with a given equilibrium size distribution, in order to minimise the overall free energy. In this work we will concentrate on the latter.

There are various candidate mechanisms for interactions between somewhat large protein-like inclusions surrounded by smaller lipid species. They range from direct, specific interactions between the inclusions, such as via van der Waals or electrostatic forces \cite{honig86_ARBBC,Lee03_BBA}, to indirect interactions mediated by the lipid membranes. For the latter, one possible mechanism arises from minimising membrane deformations. Both theory and experiments have demonstrated that, by inducing curvature, proteins and colloidal particles can self-assemble into complex structures such as lines, rings and lattices \cite{reynwar07_Nature,briggs07_PNAS,saric13_SoftMAtt,vahid16_PRL}. More recently, ideas for a new class of indirect mechanisms have emerged, namely Casimir-like forces \cite{machta12_PRL,Kardar13, Fournier11, Podgornik16, Bitbol10, golestanian96_EPL, yolcu11_EPL,hsiangKu11_PRL,weikl01_EPL}, which could arise both from shape and composition fluctuations of the membrane. Here we will focus on composition fluctuations. Initially envisioned for bulk critical mixtures in three dimensions\cite{fisher78_CRASPB}, Casimir-like forces are expected to arise between any inclusions embedded in a fluctuating fluid when separated by a distance shorter than the correlation length of the fluid; and for fluid mixtures near a critical point, this correlation length diverges as we approach the critical temperature. Interestingly, close to room temperature, lipid membranes with critical composition pass through a miscibility critical point whose nature closely follows that of the two-dimensional Ising universality class \cite{HonerkampSmith08_BioJ}. This behaviour has  been observed in giant plasma membrane vesicles isolated directly from living cells \cite{veatch08_ACSCB}, as well as in synthetic lipid mixtures \cite{connell13_Faraday}. It thus becomes plausible that Critical Casimir (CC) interactions between membrane proteins embedded in critical lipid mixtures could pave ways for their aggregation.

 Machta {\it et al.} \cite{machta12_PRL} recently demonstrated that two ``like'' inclusions immersed in a 2-dimensional critical binary lipid mixture have an attractive interaction, while ``unlike'' ones have a repulsive interaction. In their model each protein inclusion has a set uniform affinity with one of the lipid components. These findings are in line with those previously reported both theoretically and experimentally for CC interactions in 3 dimensions \cite{Krech97, hertlein08_Nature, Gambassi09_PRE, Dietrich09, nellen09_EPL}. Recent extensive simulations of colloids immersed in a critical solvent further suggest strong similarities between the 2 and 3-dimensional cases whereby the observed variations in the phase behaviour appear to be accountable to differences in critical exponent and dimension of the solvent-inclusion interface \cite{edison15_PRL, Tasio17}. 

Besides uniform inclusions, inhomogeneities and anisotropies at the fluid-inclusion interface have been theoretically shown to affect substantially CC interactions in 3 dimensions \cite{Dietrich09, toldin13_PRE, nowakowski14_JCP, labbelaurent16_SM} to the extent that patterning can now be tuned and exploited for targeted colloidal self-assembly \cite{nguyen16, labbelaurent16_SM}. This observation is particularly interesting and relevant for CC forces between membrane proteins as proteins often have multiple domains with varying affinities with respect to the lipid species\cite{cohen95_Cell,pawson03_Science,mayer15_NatRevMolCellBiol}. Thus, if CC interactions are to play a role in membrane protein aggregations, it is key to understand how their sign and magnitude can be tailored by heterogeneities in the protein domains and different affinities between the inclusions and the lipid species.

To address this issue, in this work we employ Monte Carlo (MC) simulations in a square lattice to simulate a binary mixture of lipids with two inclusions embedded in it. The free energy corresponding to the CC interactions between these inclusions is explicitly computed by using a thermodynamic integration scheme. The flexibility of our scheme allows us to (i) explore the effect of gradually modifying the inclusion boundary conditions, and (ii) study how the presence of different domains (anisotropy) can affect the interactions.

\section{Methods}

The Critical Casimir (CC) interaction is characterised by the free energy $F_{CC}(d)$ of a critical fluid mixture within which lie two inclusions separated by a fixed distance $d$ \cite{fisher78_CRASPB}. There are two common approaches to compute this CC free energy. Firstly, the free energy can be defined as minus the work done by the ensemble average of the mechanical force exerted by the fluid on the inclusions when brought from infinity to a finite separation $d$ \cite{frenkelsmit, machta12_PRL}. The equilibrium fluctuations of this mechanical force also inform experiments relying on mechanical probes to evidence CC interactions; however, some care must be taken when interpreting these fluctuations, depending on how the inclusions interact with the fluid mixtures \cite{Fournier02, Fournier11}.

Secondly, from an equilibrium statistical thermodynamics viewpoint, the statistics of configurations will be governed by a Boltzmann weight that depends on $F_{CC}(d)$. Thus, a fruitful strategy is to compute the probability weights for inclusions at varying distances, which are readily available from simulations, and invert them to recover the CC free energy as a function of distance \cite{edison15_PRL}. 

However, both the force and relative probability approaches usually need to be supplemented by a theoretical estimate at a reference distance because 2D CC interactions are very long range and it can be impractical to extend the calculations to get an effectively zero interaction at large distances \cite{machta12_PRL, edison15_PRL}. In this work, we devise a non-distance-based thermodynamic integration scheme precisely to avoid the need for a reference point, which can be difficult to obtain for complex inclusion geometries. We will detail our approach in the following subsections.

\subsection{The Model}
\label{sec:model}
Our system consists of a critical lipid binary mixture in which two proteins are embedded. We use a {\it lattice gas} model of this system with a square lattice of $N$=150x150 cells. Each cell is occupied either by a species of the binary mixture or part of a protein. This lattice size is chosen as it gives a good trade-off between reducing computational costs and finite size effects. By this we mean that up to $N=$500x500, larger lattice sizes yield results whose difference with our $N$=150x150 lattice size are not statistically significant.

\begin{figure}[ht]
%\centering
\includegraphics[width=\linewidth,keepaspectratio]{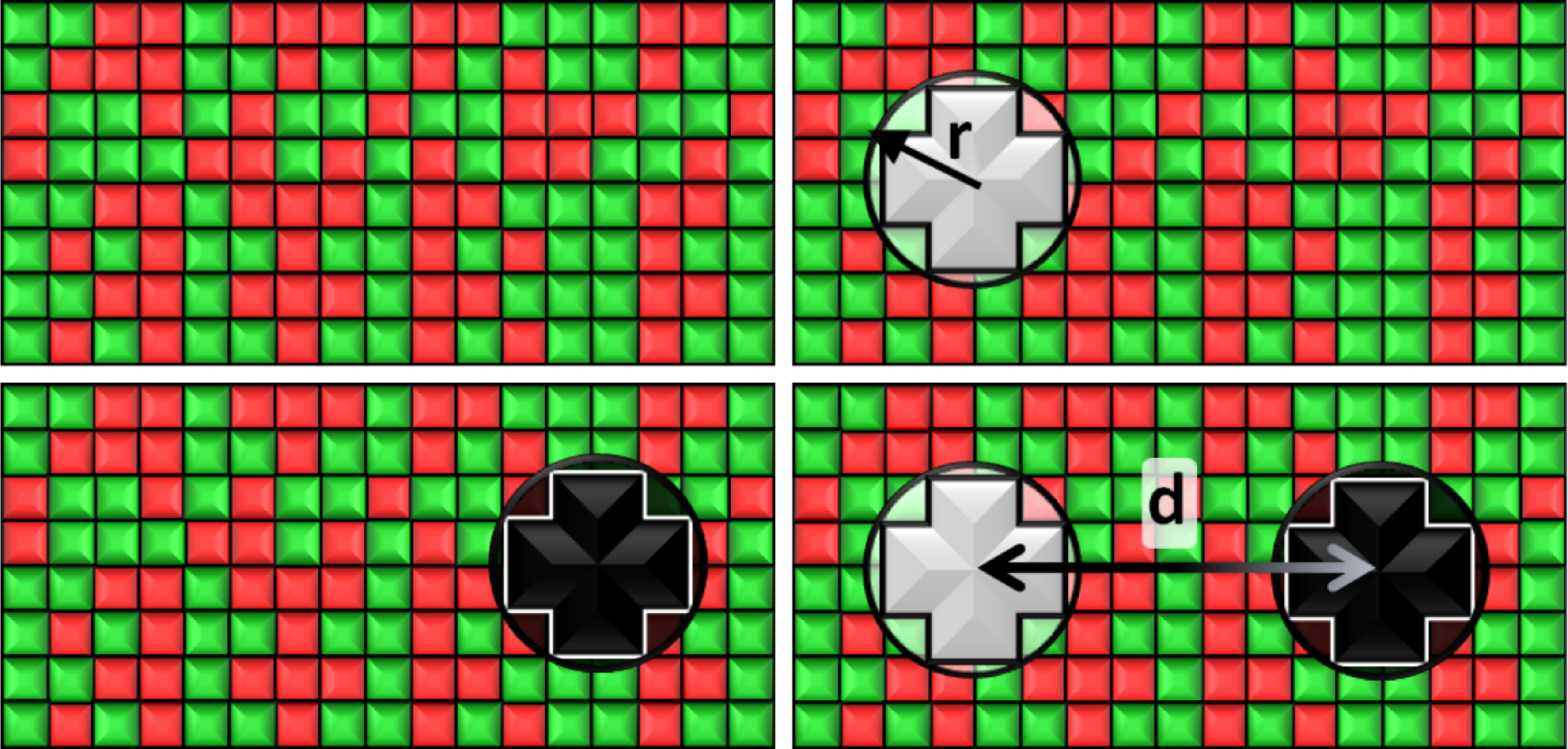} 
\caption{Representation of the various stages of the model. Top-left figure: state $\alpha$ is a binary mixture at the critical point $T_c$. Top-right figure: state $\beta$ with protein 1, alone, embedded within a critical mixture. Bottom-left figure: state $\gamma$ with protein 2, alone, embedded within a critical mixture. Bottom-right: state $\delta$ with proteins 1 and 2 embedded in a critical mixture. The proteins' effective radius, $r$, is determined by the distance 
from the center of the patch to the furthest vertex.}
\label{fig:model}
\end{figure}

We simulate the lipid mixture by making use of a two-dimensional Ising model, notorious for its critical behaviour, in which the spin variables can take values $s=\pm 1$ corresponding to lipid species A and B. The protein inclusions are modelled as two blocky patches occupying $n_p =  12$ cells each, whose size is determined by their effective radius, $r$, as shown in Fig. \ref{fig:model}. For the inclusions, the average value of their spin variable is set to a target value $s_t$ which can be different for the two proteins. To avoid a large standard deviation from this mean, spins belonging to a protein region are modelled with Potts-like spin variables $s={k/2}$, with $k$ an integer such that $k\in[-2,2]$. 

%\begin{figure}[ht]
%\centering
%\includegraphics[width=0.4\linewidth,keepaspectratio]{figures/matrix.png} \includegraphics[width=0.4\linewidth,keepaspectratio]{figures/backgroundleft.png} \\
%\includegraphics[width=0.4\linewidth,keepaspectratio]{figures/backgroundright.png} \includegraphics[width=0.4\linewidth,keepaspectratio]{figures/matrixproteins.png}
%\caption{Representation of the various stages of the model. Top-left figure: state $\alpha$ binary mixture at the critical point $T_c$. Top-right figure: state $\beta$ with protein 1, alone, embedded within a critical mixture. Bottom-left figure: state $\gamma$ with protein 2, alone, embedded within a critical mixture. Bottom-right: state $\delta$ with proteins 1 and 2 embedded in a critical mixture with the
%proteins' size determined by the distance 
%from the center of the patch to the furthest vertex.}
%\label{fig:model}
%\end{figure}

To get the free energy of the system with two protein inclusions we make use of the 4 states represented in Fig. \ref{fig:model}. The distance between proteins, $d$, is measured as the distance between their centers. These states are the following: 
\begin{itemize}
\item State $\alpha$ comprising only the binary mixture at criticality (top left of Fig. \ref{fig:model}) with hamiltonian:
\begin{equation}
H_{\alpha}=J\sum_{i=1}^{N}\sum_{\langle i \rangle} |s_i-s_j|. \label{eq:Halpha}
\end{equation}

\item State $\beta$ comprising the critical mixture and protein 1 (top right of  Fig. \ref{fig:model}) with hamiltonian:
\begin{equation}
H_{\beta}=J\sum_{i=1}^{N}\sum_{\langle i \rangle} |s_i-s_j|+h\sum_{i=1}^{n_{p1}}(s_i-s_{t1})^2. \label{eq:Hbeta}
\end{equation}

\item State $\gamma$ comprising the critical mixture and protein 2 (bottom left of  Fig. \ref{fig:model}) with hamiltonian:
\begin{equation}
H_{\gamma}=J\sum_{i=1}^{N}\sum_{\langle i \rangle} |s_i-s_j|+h\sum_{i=1}^{n_{p2}}(s_i-s_{t2})^2. \label{eq:Hgamma}
\end{equation}

\item State $\delta$ comprising the two proteins 1 and 2 separated by a distance $d/r$ within the critical mixture (bottom right of  Fig. \ref{fig:model}) and with hamiltonian:
\begin{equation}
H_{\delta}(d/r)=J\sum_{i=1}^{N}\sum_{\langle i \rangle} |s_i-s_j|+h\sum_{i=1}^{n_{p1}}(s_i-s_{t1})^2+h\sum_{i=1}^{n_{p2}}(s_i-s_{t2})^2. \label{eq:Hdelta}
\end{equation}
\end{itemize}

The first term in Eqs. \eqref{eq:Halpha}--\eqref{eq:Hdelta} is equivalent to a 2D Ising model, while the second and third terms are quadratic terms whose role is to impose the average spin values inside the regions where the proteins are. Here $J > 0$ is the coupling parameter which characterises the energy cost for having two different species next to each other, $ \sum_{\langle i \rangle}$  stands for a sum over the 4 closest neighbours of cell $i$, $h$ is a parameter setting the strength of the external potential imposing the average spin values inside the proteins, and $s_{t1}$ and $s_{t2}$ are the spin target values for proteins 1 and 2, respectively.

\subsection{Thermodynamic Integration}
\label{sec:free_ener}
Thermodynamic integration is a versatile method to compute free energy differences between two thermodynamic states. Let us consider an initial state with hamiltonian $H_{\rm ini}$, a final state with hamiltonian $H_{\rm fin}$, and a parameter-dependent hamiltonian $H(\lambda)$ such that $H(\lambda=0) = H_{\rm ini}$ and $H(\lambda=1)=H_{\rm fin}$. %Within the canonical ensemble at fixed temperature, grid size and total number of particles
It can be shown that the free energy difference between the final and initial states can be computed by \cite{frenkelsmit}:
\begin{equation}
\label{eq:ti1}
\Delta F_{\rm fin/ini} \equiv F_{\rm fin}-F_{\rm ini}=\int_0^1\left < \frac{\partial H(\lambda)}{\partial \lambda} \right >_{\lambda}d\lambda.
\end{equation}
The equilibrium averages $\left< \frac{\partial H(\lambda)}{\partial \lambda} \right >_{\lambda}$ can readily be obtained from standard MC methods. It is important to note that, as a thermodynamic quantity, the free energy difference is only a function of the final and initial states, but not the path we have taken between them. Thus, for simplicity, we have taken a linear interpolation with crossover Hamiltonian
\begin{equation}
\label{eq:ti2}
H(\lambda)=H_{\rm ini}+\lambda(H_{\rm fin}-H_{\rm ini}).
\end{equation}
This then yields 
\begin{equation}
\Delta F_{\rm fin/ini} = \int_0^1\left < H_{\rm fin}-H_{\rm ini}\right >_{\lambda}d\lambda.
\label{eq:thermo_int}
\end{equation}

Specific to the problem at hand, here we have four thermodynamic states $\alpha$, $\beta$, $\gamma$ and $\delta$ (see Fig. \ref{fig:model}), and we denote their free energies as $F_{\alpha}$, $F_{\beta}$, $F_{\gamma}$ and $F_{\delta}(d/r)$. We finally denote $F_{CC}(d/r)$ as the CC free energy and define it as the work done to bring the two proteins to within a distance $d/r$ from infinity. It follows from this definition that:
\begin{equation}
F_{CC}(d/r) = F_{\delta}(d/r)-(F_{\beta}+F_{\gamma})+F_{\alpha}, \label{eq:CCFdef}
\end{equation}where $(F_{\beta}+F_{\gamma})-F_{\alpha}$ represents the free energy of the system with the two proteins being infinitely far apart in the critical mixture. From Eq. \eqref{eq:thermo_int}, $F_{CC}(d/r)$ can then be expressed as a function of thermodynamic integrals only:
\begin{equation}
F_{CC}(d/r) = \Delta F_{\gamma/\delta} - \Delta F_{\alpha/\beta} = \int_0^1\left < H_{\delta}-H_{\gamma}\right >_{\lambda}d\lambda - \int_0^1\left < H_{\beta}-H_{\alpha}\right >_{\lambda}d\lambda \label{eq:pathway}
\end{equation}
which can be readily estimated from our MC simulations. We note that, for a given set of protein inclusions, the second thermodynamic integral only needs to be computed once, while the first integral has to be repeated for various values of $d/r$. The advantage of Eq. \eqref{eq:pathway}, which resemble to some extent the parameter variation method used in Ref. \cite{toldin13_PRE}, is that it avoids the problem of having to supply a theoretical estimate of the reference free energy, which tends to be very much system-dependent and difficult to compute analytically. %{\color{red}For example, Machta et al. overcame this shortcoming using a distance-based thermodynamic integration while avoiding very large system sizes by making use of theoretical results from Conformal Field Theory \cite{machta12_PRL}.  What does this add by saying this here? }

\subsection{Simulation Details}
We perform standard Metropolis Monte Carlo simulations at the critical temperature of the Ising model with non-conserved order parameter. In our simulations, it corresponds to $J/k_BT = 2.27$, in agreement with analytical predictions by Onsager and previous simulation results \cite{onsager44_PhysRev,mon92_JCP}. For each system we employ three random different initial configurations, which are equilibrated for $5 \times 10^5$ cycles, and sampled for $5 \times 10^6$ cycles. Each cycle consists of $N$ random single spin flips, which ensures that all particles in the system can be selected in each cycle, and configurations are saved every 1000 cycles. In order to determine the error in our calculations we use block analysis \cite{flyvbjerg89_JCP} to determine the number of independent measurements of the free energy. The error is obtained from the standard deviation with a confidence interval of 95\%. 
%block sizes = 1200 confs --> 4confs/seed --> 12 independent measurments of F

When performing the thermodynamic integration, there are two technical aspects worth commenting. The first aspect is the choice for the parameter $h$ in Eqs. \eqref{eq:Halpha}--\eqref{eq:Hdelta}. On the one hand, its value must be large enough to effectively pin the spin values inside an inclusion such that the associated energy variation $\langle H_{\rm{fin}} - H_{\rm{ini}} \rangle_{\lambda} \rightarrow 0$ as $\lambda \rightarrow 1$. On the other hand, the accuracy of our estimate of the CC free energy, $F_{CC}$, is better when we have more points substantially contributing to the numerical integration in Eq. \eqref{eq:thermo_int}. As a consequence, $h$ must be in a window of values guaranteeing both fidelity to the model we want to implement and reliability of the integration method.  

\begin{figure}[ht]
\captionsetup[subfigure]{position=t}
\centering
\includegraphics[width=\linewidth,keepaspectratio]{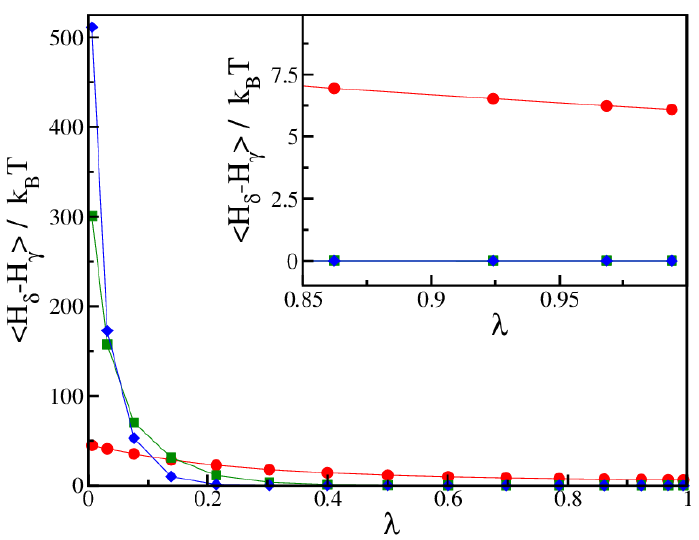}
\caption{Average values of the integrand in Eq. \ref{eq:thermo_int} as a function of $\lambda$ for different values of the external field: $h=50k_BT$ (red circles), $h=250 k_BT$ (green squares) and $h=500k_BT$ (blue diamonds). The initial state corresponds to state $\gamma$ with $s_t = -1$, and the final state corresponds to state $\delta$ with $s_t=-1$ for the first inclusion and $s_t = 0$ for the second inclusion. The distance between the two inclusions is $d/r = 2.24$. Inset: magnification in the region of high values of $\lambda$. The lines for $h=250 k_BT$ (green squares) and $h=500k_BT$ (blue diamonds) are indistinguishable.}
\label{fig:integ_value}
\end{figure}

We further illustrate this issue in Fig. \ref{fig:integ_value}, where  we show a typical result of our thermodynamic integration method for three values of $h = 50k_BT$, $250 k_BT$ and $500 k_BT$. In this example, we compute the CC free energy difference between states $\gamma$ and $\delta$. For the largest value of $h$, the contribution to the integral in Eq. \eqref{eq:thermo_int} primarily comes from a small window of $\lambda$ close to $\lambda \rightarrow 0$ and only involves a couple of integration points, leading to a poor estimate of the integral. From this perspective, smaller values of $h$ are better. However, for small values of $h$, we notice that the integrand in Eq. \eqref{eq:thermo_int} does not converge to zero as $\lambda \rightarrow 1$ (see the inset of Fig. \ref{fig:integ_value}). Physically this means the spin value of protein inclusions is not correctly set to the target value $s_t$. In this work, we find that we obtain equivalent results for the Critical Casimir free energy if we use $h = 100 - 250 \, k_BT$. For the rest of this paper, we use $h=250 k_BT$. 

\begin{figure*}[ht]
\captionsetup[subfigure]{position=t}
\centering
\includegraphics[width=0.75\linewidth,keepaspectratio]{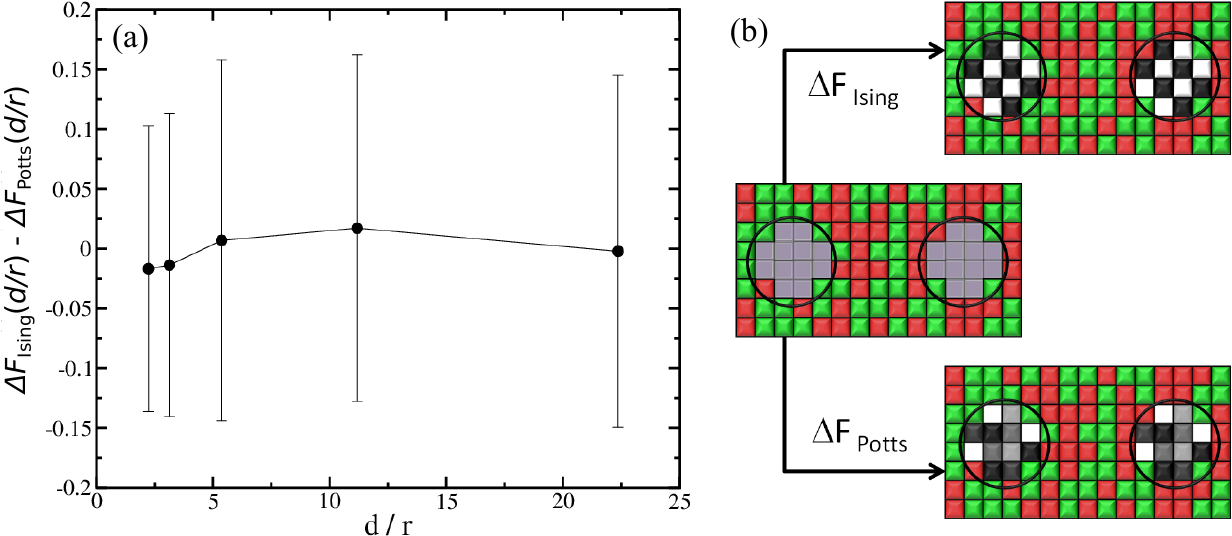}
\caption{(a) $\Delta F_{\rm Ising}(d/r)-\Delta F_{\rm Potts}(d/r)$ is the correction in the CC free energy due to variations in the spin degrees of freedom between lattice points representing lipid species and the protein inclusions, as a function of the normalised separation $d/r$. The corrections are negligible. (b) Sketch of the systems simulated. In the middle, the inclusion cells (inside the circles) do not interact with their surrounding cells or with each other. On the top right, all lattice points, including the inclusion cells, interact via standard Ising model. On the bottom right, the inclusion cells are modelled with Potts-like spin variables, while the other lattice sites have Ising spin variables.}
\label{fig:isingpotts}
\end{figure*}

The second aspect concerns the difference in degrees of freedom when a given lattice point represents a lipid species or part of a protein inclusion. For instance, when computing $\Delta F_{\alpha/\beta}$, the protein sites in state $\beta$ have five possible spin values (Potts model with $s={k/2}$, $k\in[-2,2]$), whereas the equivalent sites in state $\alpha$ only have two possible spin values (Ising model with $s =\pm 1$), because in state $\alpha$ they represent lipid species. Since we are computing the CC free energy at varying distances between the protein inclusions, in principle we must account for corrections due to variations in degrees of freedom explicitly in the computations for $F_{CC}(d/r)$. To do this, we carry out two sets of thermodynamic integration calculations. In the first set, see sketch in Fig. \ref{fig:isingpotts}(b), the initial state corresponds to the case where there is no interaction neither between the cells where the inclusions are located nor with their surrounding cells. The final state is where all cells are interacting and they all have two possible spin values (Ising model). We denote this free energy difference as $\Delta F_{\rm Ising}(d/r)$, where $d$ once again is the separation between the two inclusions and $r$ is the radius of the inclusions. For the second set of calculations, see sketch in Fig. \ref{fig:isingpotts}(b), the initial state is the same as before. However, for the final state, the cells where the inclusions are located can now have five possible spin values (Potts model). The surrounding sites still have $s =\pm 1$. The free energy difference for this case is denoted as $\Delta F_{\rm Potts}(d/r)$. The difference $\Delta F_{\rm Ising}(d/r)-\Delta F_{\rm Potts}(d/r)$ is therefore the correction in the CC free energy due to changes in the spin degrees of freedom as a function of the normalised separation $d/r$, and the typical results are shown in Fig. \ref{fig:integ_value}(a) for a 150x150 square lattice. In practise, these corrections are small and, in fact, within the uncertainty of our typical thermodynamic integration results.

\section{Results and Discussion}
We study the effective interaction between two protein inclusions embedded in a lipid membrane depending on the binding affinity that the proteins have for the lipid species A and B present in the membrane. For instance, a protein (or domain in a protein) labelled as -1 strongly prefers being surrounded by lipid species A, while a protein (or domain in a protein) labelled as 1 strongly prefers lipid species B. A value of 0 means that it has no preference for either of the species.

\subsection{Single-domain proteins}
We first illustrate the approach to the critical temperature $T_c$ by plotting two-dimensional maps of the ensemble average spin value denoted $\varphi$ for various temperatures in the neighbourhood of $T_c$. The presence of uniform inclusions with set spin value $s_t = -1$ introduces inhomogeneities into the spin system which persist over a distance of the order of the correlation length. Plotting the $\varphi-$maps enables us to have a direct visualisation of the correlation length. Indeed, the divergence of the correlation length at $T_c$ gives rise to long-range CC forces to exist.
\begin{figure}[ht]
\includegraphics[width=\linewidth]{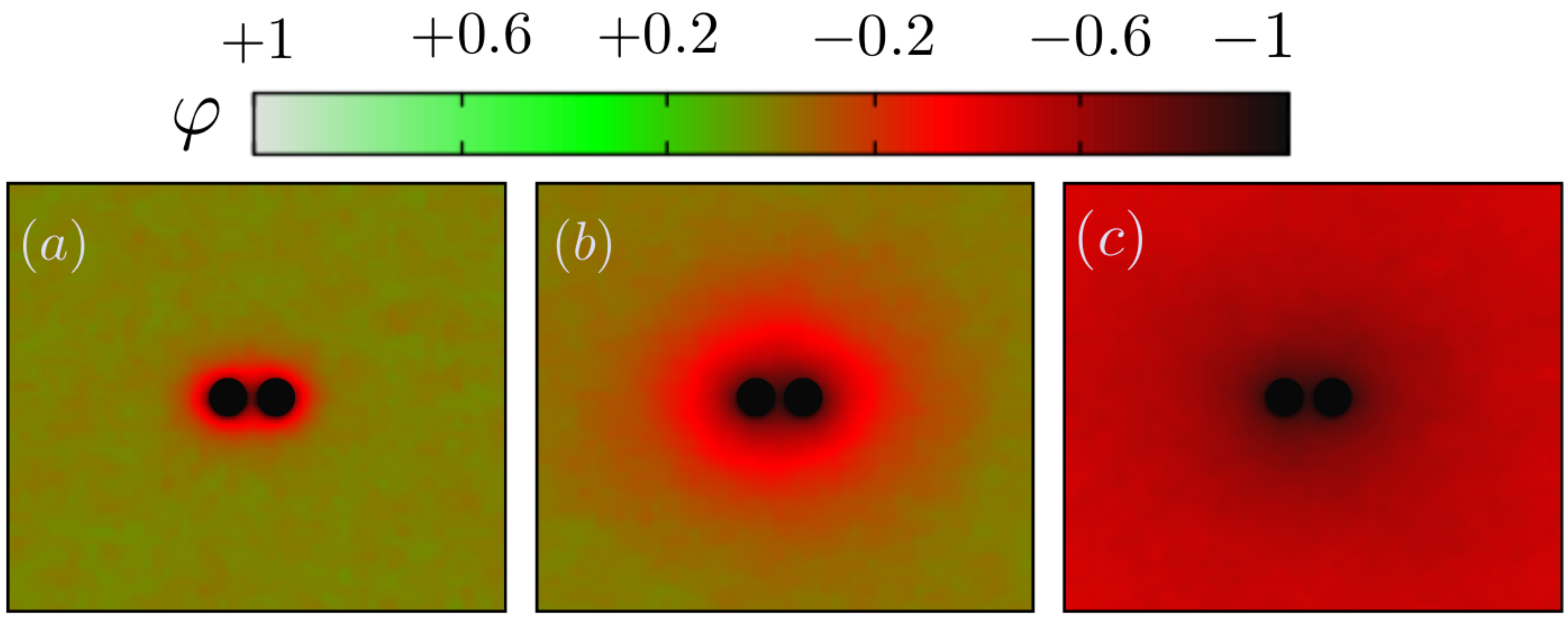}
\caption{{\it Visualising the approach to $T_c$}. 2D maps of the average spin value $\varphi$ in the vicinity of 2 inclusions (disks) with target spin value set at $-1$ (corresponding to a dark colour in the map). (a) $T = 1.33 \:T_c$, (b) $T = 1.06\:T_c$ and (c) $T = T_c$.}
\label{fig:approachTc}
\end{figure}
As shown in Fig. \ref{fig:approachTc}, approaching $T_c$ from above, the distance over which the inclusions influence the rest of the spins in the system is initially rather short (panel (a)), but increases more and more (panels (b) and (c)) until the correlation length covers the whole box at $T_c$. 

\vspace{2mm}
   
\begin{figure}[t]
\begin{subfigure}{0.5\textwidth}\centering
\includegraphics[width=\linewidth]{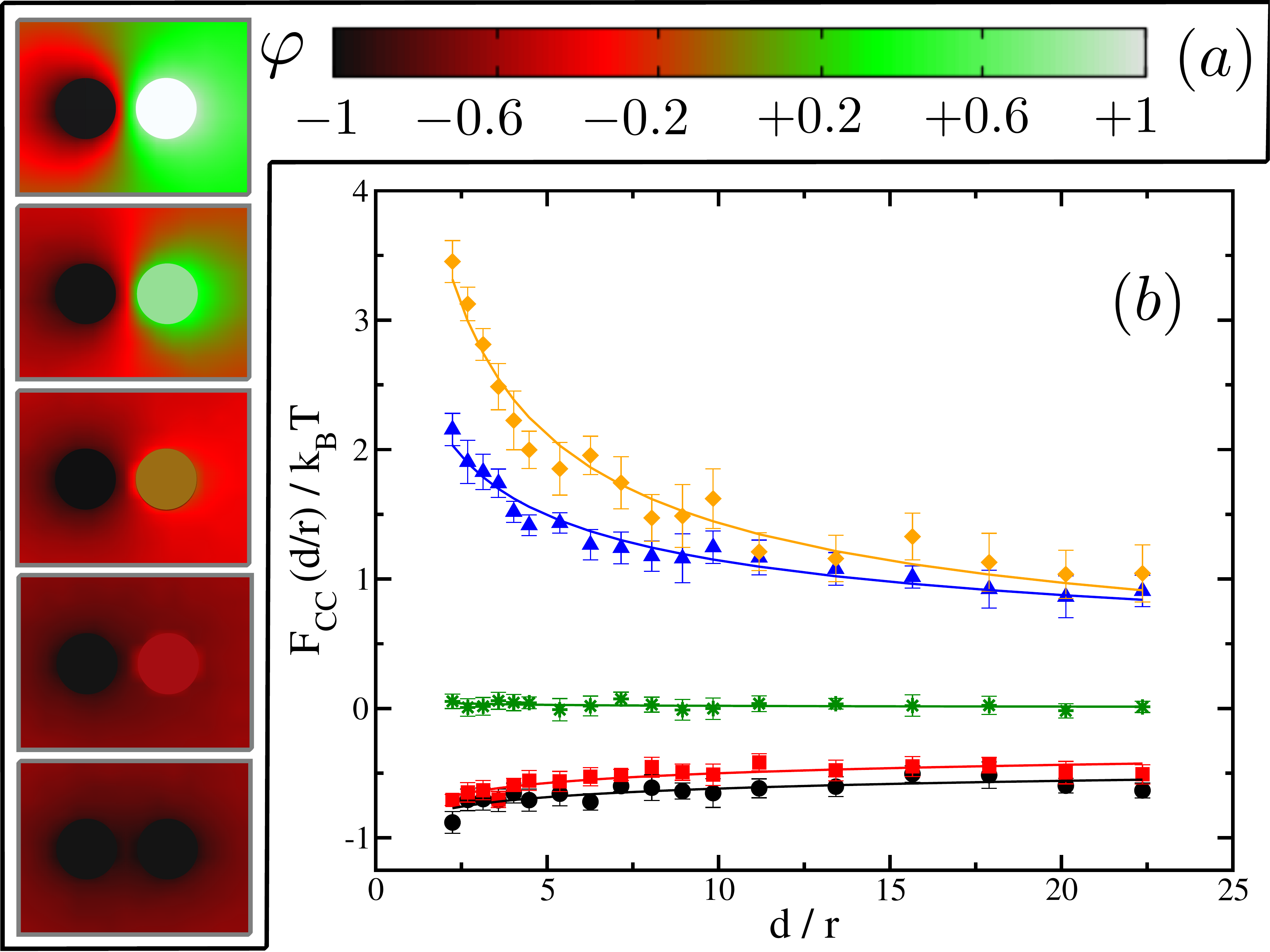}
\end{subfigure}
\begin{subfigure}{0.5\textwidth}
\small
\vspace{5mm}
%\centering 
\raisebox{8mm}{(c)} 
 \begin{tabular*}{0.9\textwidth}{cccccc}%{@{\extracolsep{\fill}}cccccc}
    \hline
    System  & (-1,1) & (-1,0.5) & (-1,0) & (-1,-0.5) & (-1,-1) \\
    \hline
    $\zeta$ & 5.2(2) & 2.8(1) & 0.06(4) &  -0.80(4) &  -0.87(5)\\
    $\nu$   & 0.56(3) & 0.38(2)  & 0.5(4) & 0.20(3) & 0.15(3) \\
    
    \hline
  \end{tabular*}
\end{subfigure}
\caption{{\it CC interactions for single-domain inclusions}. (a) From top to bottom: 2D maps of the average spin value $\varphi$ for cases $(-1,+1)$, $(-1, +0.5)$, $(-1,0)$, $(-1,- 0.5)$ and $(-1,-1)$. (b) Effective interactions as a function of distance for varying affinities of the 2nd inclusion: System $(-1,1)$ (orange diamonds), system $(-1, 0.5)$ (blue triangles) , system $(-1,0)$ (green stars), system $(-1,-0.5)$ (red squares) and system $(-1,-1)$ (black circles). (c) Fitting parameters to $F_{CC}(d/r)/k_BT = \zeta (d/r)^{-\nu}$ for the systems shown in (b).}
\label{fig:singledom_inter}
\end{figure}

We next test our thermodynamic integration scheme by computing the CC free energy for proteins that consist of one single domain. We explore the variations of the protein-protein interactions as a function of their separation for different protein binding affinities to the lipid species. In our approach, we gradually change the binding affinity of one of the proteins to the lipid species, while keeping the other one constant. More specifically, we consider systems in which protein 1 has a spin target value of -1, while protein 2 can get spin target values of 1, 0.5, 0, -0.5 or -1 (see Fig. \ref{fig:singledom_inter} for a description). The results for these systems are summarised in Fig. \ref{fig:singledom_inter}.

We start by looking at the limit in which both proteins have opposite binding affinities for the two lipid species, system $(-1,1)$. Here, protein 1 has strong preference for lipid A and protein 2 has strong preference for lipid B, which we call the "unlike" limit. In this limit the effective interaction is clearly repulsive (orange diamonds in Fig. \ref{fig:singledom_inter}) and can be quite strong with a free energy barrier rising up to $3.5 k_B T$. Then, as the binding affinity of protein 2 is gradually varied, and its preference becomes weaker for lipid species B and stronger for A, the strength of the repulsive interaction decreases (blue triangles in Fig. \ref{fig:singledom_inter}). At the point in which protein 2 has no preference for either of the lipid species, system $(-1,0)$, we find a crossover 
between the two regimes, and CC interactions become negligible with respect to $k_B T$ (green stars in Fig. \ref{fig:singledom_inter}). Finally, as protein 2 keeps on increasing its affinity for lipid A we get into the attractive regime (red squares in Fig. \ref{fig:singledom_inter}). In the "like" limit in which both proteins have the same strong affinity to one of the lipid species, system $(-1,-1)$, we find that the magnitude of attraction is the strongest with a well depth of about $1 k_B T$ (black circles in Fig. \ref{fig:singledom_inter}). However, this attraction is much weaker than the magnitude of the repulsion in the "unlike" limit, by more than a 3 fold difference. That like proteins attract supports the idea that proteins of the same kind tend to coalesce in critical lipid mixtures in order to minimise the total free energy of the system.

We note that the free energy curves in Fig. \ref{fig:singledom_inter} appear to have an offset with respect to the zero free energy baseline. However, this apparent offset is just a consequence of the very slow decay of these curves to zero as $d/r \rightarrow \infty$ and thus of the very long range nature of CC interactions in two dimensions. We further find that the free energy dependence with distance can be well fitted to a power law $F_{CC}(d/r)/k_BT = \zeta (d/r)^{-\nu}$, shown as plain lines in Fig. \ref{fig:singledom_inter}. The values for $\zeta$ and $\nu$ are tabulated in panel (c) in Fig. \ref{fig:singledom_inter}. Smaller values of $\nu$ signify longer-range of interactions. 

Our results for the like (black circles) and unlike (orange diamonds) limits in Fig. \ref{fig:singledom_inter} corroborate the findings of Machta et al. \cite{machta12_PRL} where we find excellent agreement without any fitting parameter. In their simulation work using Bennett algorithm \cite{bennett76_JCompPhys,jarzynski97_PRL}, they take advantage of asymptotic results from conformal field theory for the reference free energy at large distances. Our thermodynamic integration approach here circumvents the need for supplying the reference free energy at large distances, and thus it allows more complex scenarios to be computed readily, as we shall show in the next section for inclusions with multiple domains. Our results are also in good qualitative agreement with the experimental findings of Ref. \cite{nellen09_EPL}, where the CC forces between colloids and a wall were tuned by varying the preferential absorption properties of the wall for the critical mixture's components.

%\begin{figure}[ht]
%\centering
%\begin{subfigure}[t]{0.1\paperwidth}
%\centering
%\caption{}
%\includegraphics[width=0.95\linewidth]{figures/singledombw.png}
%\label{fig:singledom}
%\end{subfigure} 
%\begin{subfigure}[t]{0.3\paperwidth}
%\centering
%\caption{}
%\includegraphics[width=\linewidth]{figures/affinSingle.eps}
%\label{fig:singledom_inter}
%\end{subfigure} 
%\caption{(a) Sketch of the different systems for single-domain proteins. From 
%top to bottom systems: $(-1,1)$, $(-1,0.5 )$, $(-1,0 )$, $(-1, -0.5)$ and $(-1,-1)$. (b) Effective interactions as a function of distance for the different systems in (a): System $(-1,1)$ (orange diamonds), system $(-1, 0.5)$ (blue triangles) , system $(-1,0)$ (green stars), system $(-1,-0.5)$ (red squares) and system $(-1,-1)$ (black circles). {\color{red} should change label in the y axis from potential to Fcc(d)}}
%\label{fig:homo_affin}
%\end{figure}

Focusing on the attractive regime which is the strongest for like inclusions, we next look at the magnitude of the CC interactions between like proteins, as we vary their binding affinity with the surrounding matrix from strong preference for lipid A (system $(-1,-1)$) to no preference for either lipid (system $(0,0)$). Note that we do not investigate explicitly the case where the two proteins prefer lipid B since it is equivalent to the system $(-1,-1)$. Our results are summarised in Fig. \ref{fig:singledom_inter}(b). It is found that the greater the affinity of the two proteins for one of the lipid species, the greater the magnitude of the resulting CC interactions. In fact, for system (0,0) where the inclusions have no preference for either of the lipid species, we cannot measure any net interaction within the limits of our computational accuracy. For system $(-0.5,-0.5)$ we clearly observe an attractive interaction (negative free energy). However, the CC force (the slope of the free energy with distance) is very small. It is akin to the case where the system sits in an effective negative square well potential.

%\begin{figure}[ht]
%\captionsetup[subfigure]{position=t}
%\center
%\begin{subfigure}[t]{0.1\paperwidth}
%\caption{}
%\vspace{8mm}
%\raisebox{0mm}{
%\includegraphics[width=\linewidth]{figures/singledomBbw.png}}
%\label{fig:singledom2}
%\end{subfigure}
%\hfill
%\begin{subfigure}[t]{0.3\paperwidth}
%\caption{}
%\includegraphics[width=\linewidth]{figures/affindelta0.eps}
%\label{fig:singledom2_affin}
%\end{subfigure}
%\caption{(a) Sketch of different systems for single-domain proteins. From top to bottom systems: $(0,0)$, $(-0.5,-0.5 )$ and $(-1,-1)$. (b) Effective interactions as a function of distance for the different systems in (a): system $(0,0)$ (green diamonds), system $(-0.5,-0.5)$ (red squares) and system $(-1,-1)$ (black circles). {\color{red} change fit for red line}}
%\label{fig:homo_delta0}
%\end{figure}

\begin{figure}[ht]
\center
\includegraphics[width=\linewidth]{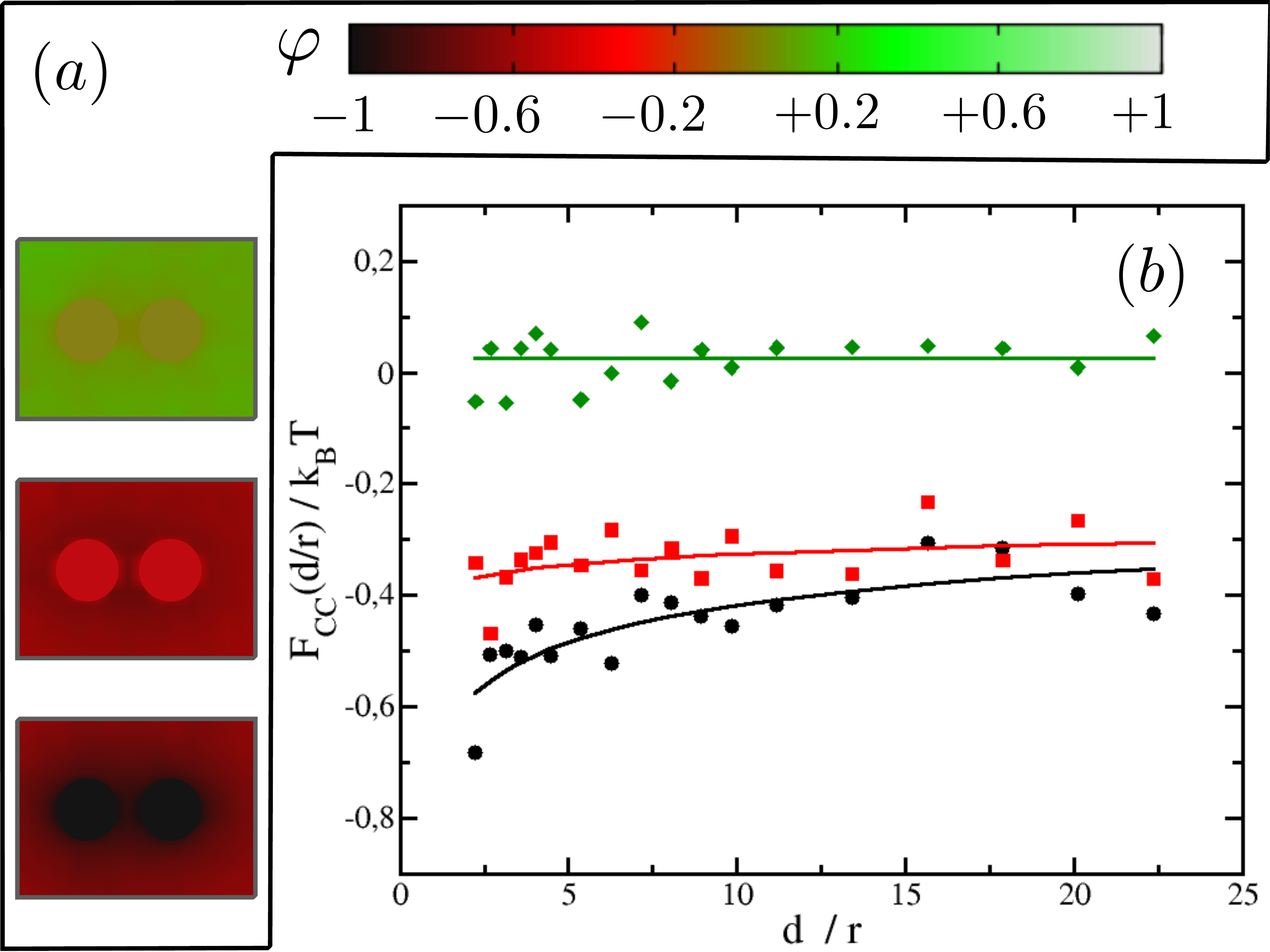}
\caption{{\it CC interactions for single-domain inclusions}. (a) From top to bottom systems: $(0,0)$, $(-0.5,-0.5 )$ and $(-1,-1)$. (b) Effective interactions as a function of distance for the different systems in (a): system $(0,0)$ (green diamonds), system $(-0.5,-0.5)$ (red squares) and system $(-1,-1)$ (black circles).}
\label{fig:homo_delta0}
\end{figure}

\subsection{Multiple-domain proteins}
In the previous section we noted the absence of CC interaction at any distance when one or both of the proteins do not have any preference for either of the lipid species. In this section we investigate whether this still remains the case for identical proteins with multiple domains each of which being given a target affinity of either -1 or 1 --- {\it i.e.} with strong binding affinity for either lipids A or B --- but such that the overall net affinity of each protein is zero. To some extent, this approach with multiple-domain proteins represents a more realistic model of proteins which usually comprise several domains with different binding affinities to the surrounding lipids \cite{cohen95_Cell,pawson03_Science,mayer15_NatRevMolCellBiol}, thus making the present study more relevant to biological systems.

%\begin{figure}[ht]
%\centering
%\begin{subfigure}[t]{0.1\paperwidth}
%\centering
%\caption{}
%\includegraphics[width=0.95\linewidth]{figures/twodomainbw.png}
%\label{fig:twodom}
%\end{subfigure} 
%\begin{subfigure}[t]{0.3\paperwidth}
%\centering
%\caption{}
%\includegraphics[width=\linewidth]{figures/affinsystem1.eps}
%\label{fig:twodom_inter}
%\end{subfigure} 
%
%\begin{subfigure}[t]{0.5\paperwidth}
%\caption{}
%\includegraphics[width=0.3\linewidth]{figures/FieldImageSystem3_try.pdf}
%\includegraphics[width=0.3\linewidth]{figures/FieldImageSystem3_try.pdf}
%\includegraphics[width=0.3\linewidth]{figures/FieldImageSystem3_try.pdf}
%\label{fig:}
%\end{subfigure} 
%\caption{(a) Sketch of the different systems for two-domain proteins. From top to bottom: system 1, system 2, system 3, system 4 and system 5. The blue domains are set with $s_t=-1$, while the orange domains have $s_t=1$. (b) Effective interactions as a function of distance for the different systems in (a): system 1 (black circles), system 2 (orange diamonds), system 3 (green stars), system 4 (red squares), and system 5 (blue triangles). {\color{red} change fit for green line}}
%\label{fig:twodom_tot}
%\end{figure}
\begin{figure}[ht]
\begin{subfigure}{0.5\textwidth}\centering
\includegraphics[width=\linewidth]{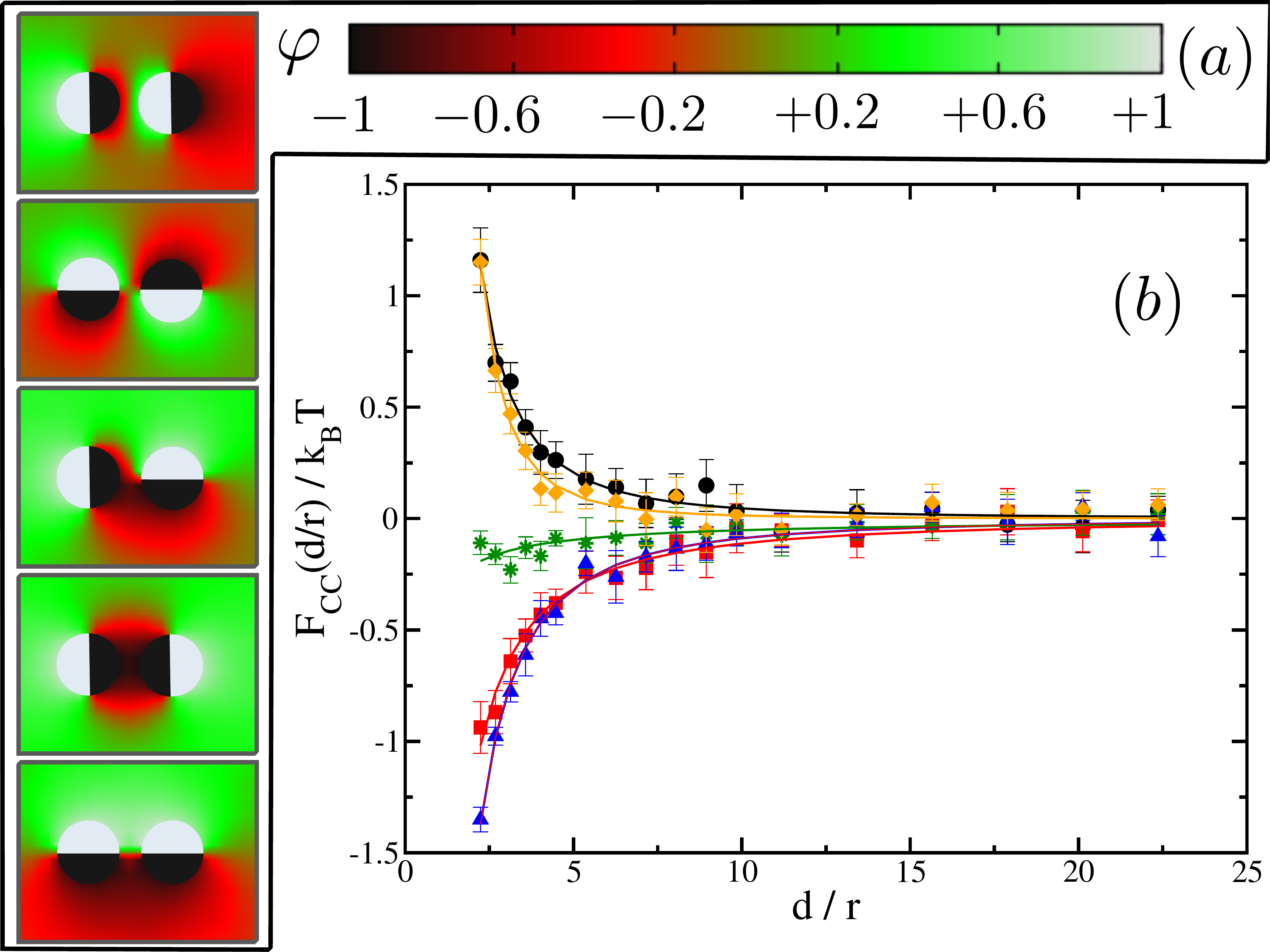}
\end{subfigure}
\begin{subfigure}{0.5\textwidth}
\small
\vspace{5mm}
%\centering 
\raisebox{8mm}{(c)} 
 \begin{tabular*}{0.9\textwidth}{cccccc}%{@{\extracolsep{\fill}}cccccc}
    \hline
    System  & 1 & 2 & 3 & 4 & 5 \\
    \hline
    $\zeta$ & 6.4(8) & 12(2) & -0.4(1) & -3.3(3)  & -6.0(5) \\
    $\nu$   & 2.2(1) & 3.0(2) & 0.9(2) & 1.7(8) & 1.83(8) \\
    
    \hline
  \end{tabular*}
\end{subfigure}
\caption{{\it CC interactions for two-domain inclusions}. (a) 2D maps of the average spin field $\varphi$ for different relative orientations of the affinity dipoles. From top to bottom: system 1, system 2, system 3, system 4 and system 5. The black domains are set with $s_t=-1$, while the white domains have $s_t=1$. (b) Effective interactions as a function of distance for the different systems in (a): system 1 (black circles), system 2 (orange diamonds), system 3 (green stars), system 4 (red squares), and system 5 (blue triangles). (c) Fitting parameters to $F_{CC}(d/r)/k_BT = \zeta (d/r)^{-\nu}$ for the systems shown in (b).}
\label{fig:twodom_inter}
\end{figure}

We start by studying interactions between two proteins with two different domains each. These proteins can have different relative orientations depending on which domains of the proteins are facing each other (see Fig. \ref{fig:twodom_inter}(a) for a description). Our aim is to study how orientation can affect these interactions and if there are preferential orientations that could lead to patterning in protein clusters. Given that these proteins do not have a net preference for any lipid type, one could naively expect no interaction between them as in the cases of the $(0,0)$ and the $(-1,0)$ pairs of single-domain proteins. However, our results as shown in Fig. \ref{fig:twodom_inter}(b) show again the presence of two different regimes and a crossover between them. We find that systems 1 and 2 appear in the repulsive regime while systems 4 and 5 fall in the attractive one. System 3, corresponding to a $90^{\rm o}$ relative orientation, is slightly attractive, though its magnitude is very small compared to the other systems. 

A closer look at systems 1 and 2 shows that, in both cases, proteins are facing each other with domains that have opposite preferences for the lipid species, giving rise to predominantly "unlike" interactions. Therefore, this situation is qualitatively analogous to the system $(-1,1)$ for single-domain proteins. On the other hand, systems 4 and 5 are confronting domains of the same kind, leading to predominantly "like" interactions as in system $(-1,-1)$ for single-domain proteins. Lastly, in system 3, proteins are orientated in such a way that half of protein 1 has "like" interactions with protein 2, while the other half has "unlike" interactions. This case is a good example of the non-additive nature of CC interactions. If they were simply additive, taking advantage of the results in Fig. \ref{fig:singledom_inter}, we would have expected the net interaction to be repulsive. The magnitude of repulsive interactions between unlike uniform inclusions is stronger than attractions between like ones. However, as shown in Fig. \ref{fig:twodom_inter}, for system 3, we observe a very weak attractive interaction instead. 

When compared with single-domain proteins, the range of the emerging CC interaction is clearly shorter in all cases, getting close to zero at distances of about ten times the radius of the protein; we interpret this as being a partial {\it screening effect} analogous to electrostatic screening in dielectric systems and owing to proteins comprising domains with opposing affinities. Panel (c) in Fig. \ref{fig:twodom_inter} tabulates the power law fit of the CC free energy with distance, $F_{CC}(d/r)/k_BT = \zeta (d/r)^{-\nu}$. Here, the exponents $\nu$ are clearly larger for the values for single-domain inclusions in Fig. \ref{fig:singledom_inter}. 
%However, as shown in Fig. \ref{fig:twodom_inter}, interactions can vary again 
%from repulsive to attractive depending on the orientation, but their range 
%is shorter than the one obtained with single domain proteins. 
%Both systems 1 and 5 present repulsive interactions while those of systems 2 
%and 4 are attractive, system 3 doesn't provide any kind of interactions at 
%any distance.  A careful look at systems 1 and 5 shows that in these systems 
%proteins are facing each other with domains that have the opposite preference 
%for the lipid specie, with great resemblance of the system 0-1 for single 
%domain proteins. On the other hand in systems 2 and 4 proteins are confronting 
%domains with the same preferance in a way very similar to that of the system 
%0-0 for single domain proteins. 

\begin{figure}[t]
\begin{subfigure}{0.5\textwidth}\centering
\includegraphics[width=\linewidth]{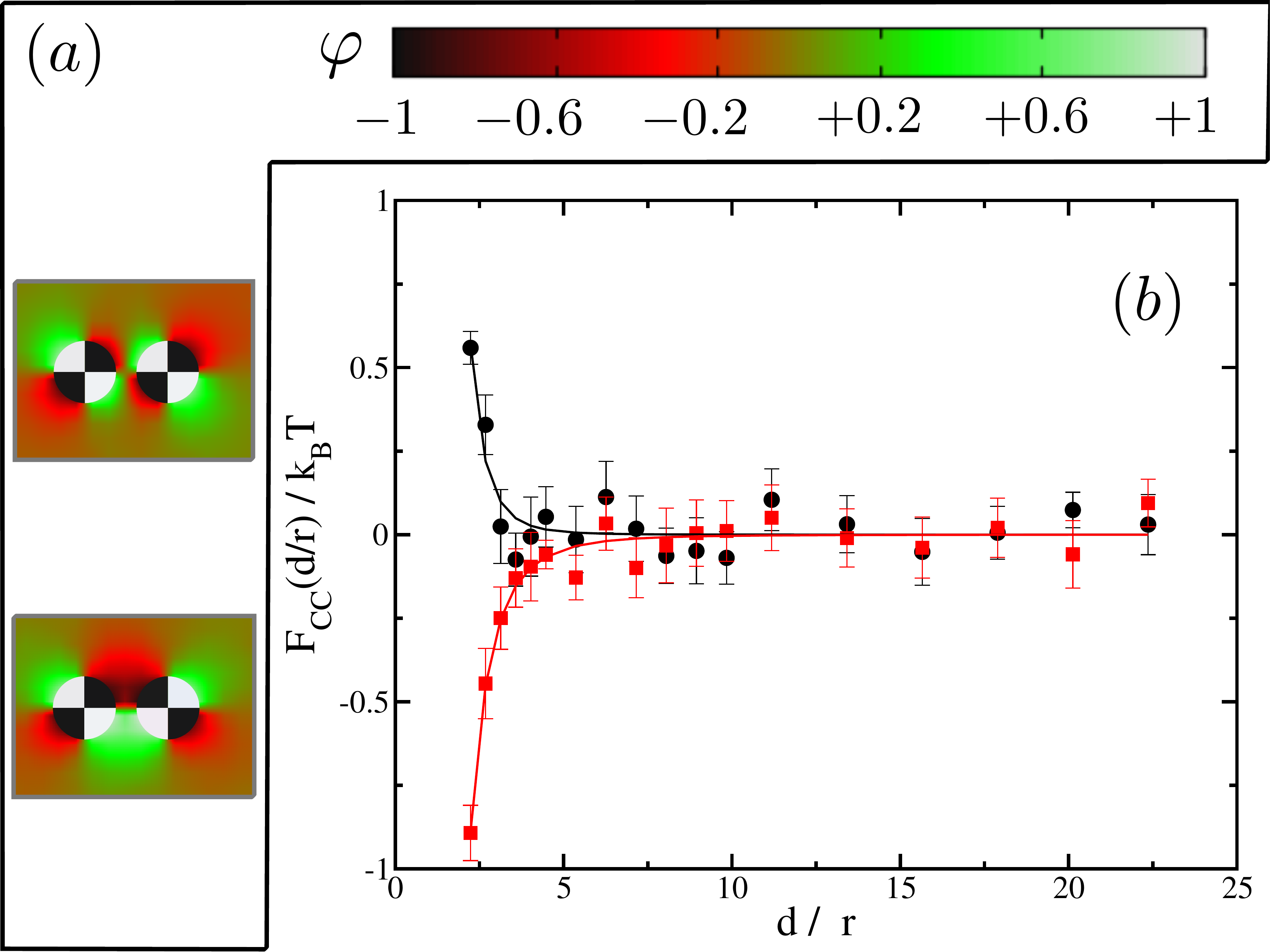}
\end{subfigure}
\begin{subfigure}{0.5\textwidth}
\small
\vspace{5mm}
\centering
\raisebox{8mm}{(c)}
 \begin{tabular*}{0.5\textwidth}{@{\extracolsep{\fill}}ccc}
    \hline
    System & 6 & 7 \\
    \hline
    $\zeta$& 50(50) & -18(5) \\
    $\nu$  & 5(1) & 3.7(3) \\
    \hline
  \end{tabular*}
\end{subfigure}

\caption{{\it CC interactions between four-domain inclusions}. (a) 2D maps of the average spin value $\varphi$ for two different orientations of the affinity quadrupoles. From top to bottom: system 6, system 7. The black domains are set with $s_t=-1$, while the white domains have $s_t=1$. (b) Effective interactions as a function of distance for the different systems in (a): system 6 (black circles), system 7 (red squares). (c) Fitting parameters to $F_{CC}(d/r)/k_BT = \zeta (d/r)^{-\nu}$ for the systems shown in (b).}
\label{fig:fourdom_inter}
\end{figure}

In order to further investigate the effect of the domains we next turn to proteins with four distinct domains. In this case only two possible orientations are of interest, see Fig. \ref{fig:fourdom_inter}(a). We present our results in Fig. \ref{fig:fourdom_inter}(b). Again, in spite of having no net preference for either of the species, proteins show either attractive or repulsive interactions. A similar analysis to that done with two-domain proteins show that system 6 confronts domains with opposite preferences for the lipid species, falling, therefore, in the repulsive regime. On the other hand system 7 has domains with the same preference facing each other, leading to attractive interactions. It is worth pointing out that, in these cases, both the range and the magnitude of the interactions are smaller than those observed for single- or two-domain inclusions. Fig. \ref{fig:fourdom_inter}(c) tabulates the power law fit for the CC free energy with distance between two four-domain inclusions. Eventually in the limit of the domain size going to zero, we expect to recover the uniform case of $s_t = 0$. Comparing the results in Figs. \ref{fig:singledom_inter}, \ref{fig:twodom_inter} and \ref{fig:fourdom_inter}, we also observe that the magnitude of the repulsive interactions falls faster than attraction interactions with increasing number of domains. 

The observed behaviours for multiple-domain inclusions are in good qualitative agreement with the findings of Toldin et al. for CC forces between chemically striped surfaces immersed in critical liquid mixtures \cite{toldin13_PRE}. In their study, they confront a surface with a strong preference for one component of the binary liquid mixtures with a striped surface in which the preference of the stripes for the liquid components is alternating. They observe a crossover between a regime with noticeable CC interaction and one with no noticeable CC interaction (similar to our $(0,0)$ and $(-1,0)$ cases for single-domain proteins) as the strip width is decreased from large values to zero.

Finally, we note that the range and strength of the CC interactions depend only on the lattice points which are at the boundary of the inclusions, and not the inner ones. This is as expected because we only have nearest neighbour interactions, and thus the inner sites do not interact directly with the surrounding cells representing the lipid mixtures.

%Casimir forces depend only on the boundary conditions of the proteins as in the case of planar surfaces immersed in a critical matrix \cite{rafai07_PhysA}.
%\begin{figure}[ht]
%\centering
%\begin{subfigure}[t]{0.1\paperwidth}
%\centering
%\caption{}
%\vspace{13mm}
%\includegraphics[width=0.95\linewidth]{figures/fourdombw.png}
%\label{fig:fourdom}
%\end{subfigure} 
%\begin{subfigure}[t]{0.3\paperwidth}
%\centering
%\caption{}
%\includegraphics[width=\linewidth]{figures/affinsystem2.eps}
%\label{fig:fourdom_inter}
%\end{subfigure} 
%\caption{(a) Sketch of the different systems for four-domain proteins. From top to bottom: system 6, system 7. The blue domains are set with $s_t=-1$, while the orange domains have $s_t=1$. (b) Effective interactions as a function of distance for the different systems in (a): system 6 (black circles), system 7 (red squares). {\color{red} missing d/r in the x-axis.}}
%\label{fig:fourdom_tot}
%\end{figure}

%\begin{figure}[ht]
%\centering
%\label{fig:fourdom_inter}
%\end{figure}

\section{Conclusions and Future Work}
In this work, we have numerically computed the CC free energy between inclusions in two-dimensional critical binary mixtures. To this end, we have introduced a versatile thermodynamic integration scheme whose calculations for single-domain inclusions compare very favourably with the existing literature on the subject\cite{machta12_PRL}. This has enabled us to identify repulsive and attractive CC regimes respectively for unlike inclusions --- with opposing affinities for the surrounding species comprising the binary mixture --- or like inclusions --- with the same affinity for the surrounding species. We noted that the CC interactions for single-domain inclusions are very long range and their magnitude is increasing with the strength of the affinity of the domain with either of the surrounding species. 

A key advantage of our thermodynamic integration approach is that we do not need to supply a reference free energy at large distances, which can be notoriously difficult to compute analytically for complex scenarios. In turn this allows us to study inclusions comprising multiple domains such that, overall, they do not have any net preference for either of the surrounding species in the mixture. 

The presence of domains with opposing affinities does not cancel the CC interactions altogether but instead leads to a decrease in the range and magnitude of the effective interactions in a manner reminiscent of electrostatic multipoles. For such multiple-domain inclusions, both repulsive and attractive CC interactions were observed depending on their relative orientations. 

If biologically relevant lipid membranes are close to criticality as suggested by some studies \cite{HonerkampSmith08_BioJ,veatch08_ACSCB}, then these findings may prove useful to better understand lateral spatial organisation within cellular membranes. As a matter of fact, attractive CC interactions between the single-domain inclusions of the same kind provide both sufficient magnitude and range for protein clustering to occur in equilibrium conditions\cite{edison15_PRL,nguyen16}. Utilising multiple-domain inclusions as more realistic models of membrane proteins lead us to the conclusion that the emerging CC interactions between them also suffice for them to aggregate, even where there is no overall net affinity to the surrounding species in the mixture.

\begin{figure}[ht]
\centering
\includegraphics[width=1.0\linewidth]{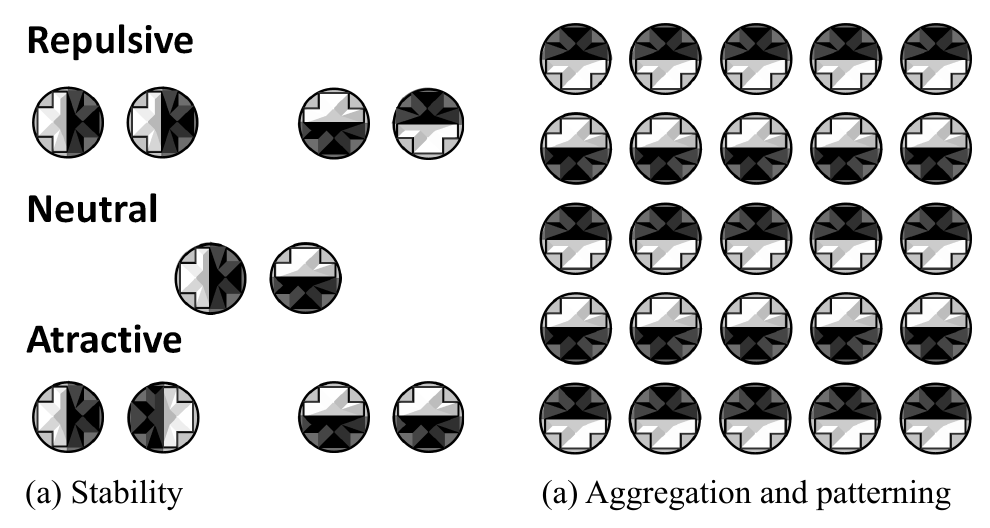}
\caption{Stability, aggregation and patterning for two-domain proteins.}
\label{fig:rot&ag}
\end{figure}

There are several avenues for future work. Firstly, it would be interesting to study the clustering, and more generally the phase diagram, of the multiple-domain inclusions. For instance, in Fig. \ref{fig:rot&ag}(a) we show the relative stability of the different orientations for two-domain proteins. Since some orientations are energetically favourable, we would expect proteins to rotate in order to minimise the overall free energy of the system. Our preliminary results suggest that, as these inclusions start to cluster, they will form an alternating patterning in affinities with respect to the surrounding species (Fig. \ref{fig:rot&ag}(b)). In this context, it would be exciting to explore how CC interactions could be tailored to allow components in biological membranes achieve specificity as suggested in Ref. \cite{sieber06_BioJour}. Such strategies could also be exploited for the self-assembly of anisotropic (e.g. Janus) colloidal particles \cite{labbelaurent16_SM,noruzifar09_PRE}. Secondly, Casimir interaction due to shape fluctuations of the membrane has also been proposed as a non-specific mechanism for membrane proteins to interact \cite{golestanian96_EPL, Bitbol10, yolcu11_EPL,hsiangKu11_PRL,weikl01_EPL}.  Here it will be instructive to compare the strengths of the two possible Casimir interactions, and map regions in parameter space (temperature, composition and bending rigidity) where one mechanism dominates over the other, and where they interfere either constructively or destructively. For instance, Dean et al.\cite{Dean15} suggests that taking into account membrane shape fluctuations can result in a shift in the critical temperature at which phase separation occurs for the lipid mixtures. Stress tensor-based methods following Ref. \cite{Bitbol11_EPJE} could help inform on these competing effects.

%In the same trend as in Ref \cite{sieber06_BioJour} we don't expect this to be 
%the main mechanism for protein aggregation given that the magnitude of the 
%attractive forces is not too big. However it can be regarded as a  
%subtle refining mechanism in which membrane protein taylor their interactions 
%in order to adquire especifity.  

%With this work we have shown that multi-domains proteins present preferable 
%orientations that reduce the energy of the system. We expect membrane 
%proteins to rotate in order to get to this preferent orientations, providing 
%a mechanism to increase the specifity of the proteins. Besides, we expect 
%this mechanism to be behind the patterning present in protein clusters, which 
%can be very helpful for medical community. 

\section*{Conflicts of interest}
There are no conflicts of interest to declare.

\section*{Acknowledments}
We thank Pietro Cicuta, Mark Miller and Lorenzo di Michele for helpful discussions. We acknowledge funding from EPSRC, grant EP/J017566/1.

%%%END OF MAIN TEXT%%%


%\balance


%%%REFERENCES%%%
\bibliographystyle{apsrev4-1}
\bibliography{biblio_inclusions} %You need to replace "rsc" on this line with the name of your .bib file

\end{document}